\documentstyle[twocolumn,prl,aps,epsfig]{revtex}
\newcommand{\beq}{\begin{eqnarray}} 
\newcommand{\eeq}{\end{eqnarray}} 
\renewcommand{\vec}[1]{{\mathbf{#1}}}
\begin{document} 
\draft 
 
\title 
{Fluctuation Conductivity in Insulator-Superconductor Transitions with Dissipation}
\author{Denis Dalidovich and Philip Phillips}\vspace{.05in}

%
\address
{Loomis Laboratory of Physics\\
University of Illinois at Urbana-Champaign\\
1100 W.Green St., Urbana, IL, 61801-3080}

%

\address{\mbox{ }}
\address{\parbox{14.5cm}{\rm \mbox{ }\mbox{ }
We analyze the fluctuation conductivity near the
critical point
in a 2D Josephson junction array shunted
by an Ohmic resistor.  We 
find that at the Gaussian level, the conductivity acquires
a logarithmic dependence on
$T/m$ ($m$, the inverse correlation length) when the dissipation is 
sufficiently weak. 
In the renormalized classical regime, this logarithmic dependence
gives rise to a leveling of the resistivity at low to intermediate
temperatures when
fluctuations are included. 
We show, however, that this trend does not
persist to $T=0$ at which point the resistivity vanishes.
}}
\address{\mbox{ }}
\address{\mbox{ }}

\maketitle

In two dimensional superconductors, the resistivity vanishes not at the
mean-field temperature, $T_{\rm BCS}$, at which the pair amplitude is
established but at a lower temperature, $T_c$, where global
phase coherence obtains\cite{review}.  The prominence of phase fluctuations in 2D 
is primarily responsible for the discrepancy between the onset of 
pairing and the
subsequent thermodynamic phase transition to a state with zero resistance.
In 2D the zero-resistance state posseses algebraic long-range
order~\cite{bez,kt,doniach}.  Above $T_c$ but below $T_{\rm BCS}$,
the system is in a paracoherent state characterised by incoherent motion of the
Cooper pairs.  In this regime, phase fluctuations give rise to the  
conductivity which at the mean-field level is
of the Aslamazov-Larkin form~\cite{al}. 

We are concerned in this work with the role of dissipation on the conductivity 
in the regime where phase fluctuations dominate.  
While thermally-excited vortices are the most common dissipative mechanism 
in 2D superconductors in the presence of a
magnetic field, we specialize to the zero field case
and consider here the case of shunting an array of
Josephson junctions with
an Ohmic resistor\cite{cl}.  Resistively-shunted Josephson junction 
arrays\cite{chak,fisher,zwerg,chak2,ks,amb,esa,wag} in 2D 
represent an idealization of an important class of insulator-superconductor 
systems,
namely granular superconductors\cite{goldman,mooij,paal} in which the 
role of Coulomb interactions and dissipation is paramount. At
$T=0$, such systems undergo an insulator-superconductor transition (IST) when the 
Josephson coupling energy exceeds the Coulomb charging energy.  Dissipation
suppresses quantum fluctuations and hence favours the superconducting state.
However, in the paracoherent regime, dissipation can 
also slow the motion of Cooper pairs. Naive considerations suggest that the 
fluctuation conductivity should decrease by an amount proportional to the 
inverse of the strength of the dissipation. In fact, 
Mason and Kapitulnik~\cite{mk}
have recently suggested that dissipation in the presence
of disorder can give rise to a leveling-off of the
resistivity at $T=0$--that is, a ``metallic state''. 

We calculate here the role dissipation and fluctuation effects play
in the fluctuation conductivity in 2D Josephson junction arrays.
Within a Gaussian theory, we show that the standard\cite{zwerg}
 form for the fluctuation
conductivity does not hold when the strength of the dissipation is
less than the gap proportional to the
inverse correlation length.  In this regime, we find that quite generally
the conductivity scales logarithmically with the strength of the gap.  While calculations of the 
fluctuation conductivity have been performed previously\cite{zwerg}, none have revealed
the regime with the logarithmic behavior at finite temperature.
We then analyze the effect of the quartic term on the fluctuation
conductivity in the $N=\infty$ limit. In a strictly $D=2$ system, the
$N\rightarrow\infty$ limit results in ordering only at $T=0$\cite{chubukov}. 
 We show that in the renormalized classical regime, the logarithmic
temperature dependence is transformed into a constant resistivity
at low to intermediate temperatures.  However, at sufficiently low temperatures,
the resistivity is shown to vanish exponentially.  

To formulate the conductivity, we replace the microscopic Hamiltonian
for an array of Josephson junctions with an effective Ginsburg-Landau
theory. The coarse-graining approximation of 
Doniach\cite{doniach} offers a straightforward way of obtaining
the appropriate Landau theory.  To this end, we introduce the complex
order parameter $\psi(\vec r,\tau)$ whose expectation value
is proportional to $\langle\exp(i\phi)\rangle$, where $\phi$ is the phase
of a particular junction.  An effective Landau theory is valid if $\psi$ is sufficiently
small as is the case in the vicinity of the onset of global phase coherence.
The minimal Ginsburg-Landau 
theory\cite{chak,fisher,zwerg,chak2,ks,amb,esa,otterlo} required to model 
quantum fluctuations and dissipation near the zero-resistance quantum 
critical point is the Gaussian free energy functional,
\beq\label{gaussian}
F[\psi]&=&\int d^2r\int d\tau\left\{
\left[\left(\nabla+\frac{ie^*}{\hbar}\vec A(\vec r,\tau)\right)
\psi^*(\vec r,\tau)\right]\right.\nonumber\\
&&\left.\cdot
\left[\left(\nabla-\frac{ie^*}{\hbar}\vec 
A(\vec r,\tau)\right)\psi(\vec r, \tau)\right]\right.\nonumber\\
&&\left.+\kappa^2\left|\partial_\tau\psi(\vec r,\tau)\right|^2
+m^2\left|\psi(\vec r,\tau)\right|^2\right\}+L_{\rm dis} 
\eeq
where $\vec A(\vec r,\tau)$ is the vector potential, $e^*=2e$, $m^2$ is proportional
to the inverse
correlation length, and $\kappa$ and $\eta$ measure the strength of the 
quantum fluctuations and dissipation, respectively. In Fourier
space, the dissipation term, 
$L_{\rm dis}=\eta\sum_{\vec k,\omega_n}|\omega_n||\psi(\vec k,\omega_n)|^2$,
corresponds to the phenomenological Ohmic model introduced by 
Caldeira and Leggett~\cite{cl}. 
We have retained the vector potential,  $\vec A(\vec r,\tau)$,
to facilitate the calculation of the conductivity. 
Along the imaginary frequency axis,
the conductivity 
\beq
\sigma_{\alpha\beta}(i\omega_n,\vec q)=-\frac{\hbar}{\omega_n}\int d^2r\int d\tau
\frac{\delta^2\ln Z}{\delta A_\alpha(\tau,\bf r)\delta A_\beta(0)}
e^{i\bf q\cdot\bf r+i\omega_n\tau}\nonumber
\eeq
is related to the fourier transform of the
 second variation of the partition function $Z$ 
with respect to the vector potential. The partition function is defined in 
the usual way as the functional integral of $\exp(-F[\psi])$. 

In the zero wavevector limit, the conductivity 
to 1-loop order\cite{otterlo} can be rewritten as
\beq
\sigma(i\omega_n)&=&\frac{2(e^*)^2}{\hbar\omega_n}T\sum_{\omega_n}
\int \frac{d^2k}{(2\pi)^2}2k_{x}^2 G(\vec k,\omega_m)\nonumber\\
&&\left(G(\vec k, \omega_m)-G(\vec k,\omega_m+\omega_n)\right),
\eeq
where $G(\vec k, \omega_n)$ is the standard Matsubara Green function.
To obtain the conductivity for real frequencies, we must analytically continue
to real frequencies.  As van Otterlo, et. al.~\cite{otterlo} have pointed out, inclusion of 
dissipation changes the analytical properties of the Matsubara sums. 
We define the retarded and advanced Green functions, 
$G^R(z)=(k^2+m^2-\kappa^2z^2-i\eta z)^{-1}$ and 
$ G^A(z)=(k^2+m^2-\kappa^2z^2+i\eta z)^{-1}$. 
Upon choosing the appropriate contour, we find that 
\beq
\sigma(\omega)&=&\frac{(e^*)^2}{2\pi\hbar\omega}
\int_0^\infty k^3 dk\int_{-\infty}^{\infty} 
\coth\frac{z}{2T} dz\left[(G^R(z)-G^A(z))\right.\nonumber\\
&&\left.[G^R(z)+G^A(z)-G^R(z+\omega)-G^A(z-\omega)]\right].
\eeq
We are particularly interested in the limit of zero frequency.  In this limit,
the product of Green functions simplifies significantly leading to
\beq\label{weq}
\sigma(\omega=0)&=&\frac{(e^*)^2}{2\pi h}\int_0^\infty k^3 
dk\int_{-\infty}^{\infty}\frac{dx}{\sinh^2 x}\nonumber\\
&&\times\frac{8\eta^2T^2x^2}{\left[(\epsilon_k^2-4T^2
\kappa^2x^2)^2+4T^2\eta^2x^2\right]^2}
\eeq
as our working expression for the zero-frequency conductivity. In this 
expression, $\epsilon^2_k=k^2+m^2$.

From the analytical structure of the integrand in Eq.~(\ref{weq}), it is 
clear that distinctly different regimes arise when 1) $m\ll\kappa T$ 
and 2) $m\gg\kappa T$. In each of these cases, we explore the limiting 
form for the conductivity as the magnitude of the dissipation varies. 
In our treatment, we explicitly assume that $\eta/\kappa\ll 1$. 
In the limit of $\eta=0$, the pole in Eq.~(\ref{weq}) makes 
the conductivity infinite at least to 1-loop
order as obtained previously by 
van Otterlo, et. al.~\cite{otterlo} and Damle and Sachdev\cite{ds}. 
Two loop and higher order corrections regularize the
static conductivity even when $\eta=0$, however.
Nonetheless, 
as the dissipation is turned on,
the divergent 1-loop conductivity becomes finite. In the
present treatment, we address only
the form of the static conductivity at lowest order.  We focus first
on the region characterized by $m\ll\kappa T$. 
The two subregimes 
of interest are  
a)  $\eta/\kappa\ll m$ and
b) $m\ll\eta/\kappa$.  Only the latter regime has been considered
previously~\cite{zwerg}.  It is in the former regime that the 
new logarithmic behaviour emerges.  To obtain this result, we note that
the main contribution from the integral in Eq.~(\ref{weq}) arises from
small $x$ and the minima of the denominator of the integrand which occur
at $x_{\pm}^2=1/4\kappa^2T^2[\epsilon_k^2-\eta^2/2T^2]$.  The 
latter two roots are real only if $m>\eta/\kappa$.  If we work in the limit
in which $m\gg\eta/\kappa$, we expand the integral about $x_\pm$ and perform
the $x$ integration
\beq\label{ultra}
\sigma=\frac{2e^2\kappa^2T}{h\eta}\int_0^\infty\frac{\xi^3d\xi}
{(\xi^2+\bar{m}^2)\sinh^2\sqrt{\xi^2+\bar{m}^2}}
\eeq
by introducing the new variables, $\xi=k/2\kappa T$ and
$\bar{m}=m/2\kappa T$. In the limit that $\bar{m}\ll 1$, we can 
approximate $\sinh^2\sqrt{\xi^2+\bar{m}^2}$ as $\xi^2+\bar{m}^2$.  This 
approximation is valid if $\xi^2\ll 1$ or equivalently an ultraviolet
cutoff is introduced on the momentum of order $2\kappa T$.  Introducing this
cutoff into the integral in Eq.~(\ref{ultra}) leads immediately to the
logarithmic form for the fluctuation conductivity: 
\beq\label{eqln}
\sigma=\frac{2e^2}{\pi h}\frac{\pi\kappa^2T}{\eta}\ln\frac{\kappa T}{m}\quad \eta/\kappa\ll m.
\eeq
In the regime that $m\ll\eta/\kappa$, the original integral in the conductivity
is dominated by the minimum at $x=0$.  Performing the integrations
in this regime results in the more standard form~\cite{zwerg} for the 
conductivity:
\beq\label{eqal}
\sigma=\frac{e^2}{h}\frac{\eta T}{m^2}\quad m\ll\eta/\kappa.
\eeq
Should the transition to the superconducting state occur at the finite 
temperature $T_c$ and the effective field theory close to the transition is
given by Eq.~(\ref{gaussian}) with $m^2\propto (T-T_c)$, we find that
the fluctuation conductivity is consistent with the standard 
Aslamazov-Larkin~\cite{al}
form when the dissipation is of intermediate strength.  However,
in the weak dissipation limit, that is, weak in comparison to the 
inverse correlation length, the standard $1/(T-T_c)$ is replaced
 by the logarithimic behavior sufficiently close to $T_c$.  

While the logarithmic conductivity follows naturally from an expansion in powers
of $\eta/(\kappa m)$ in the conductivity, this appears to be the first time this
result has been derived. Consider the regime $m\gg\kappa T$. Fluctuations are
suppressed in this regime.
In this case, $x_\pm\gg 1$ and its subsequent contribution to the
 integral is exponentially small. Thus, it will compete with the contribution
from the vicinity of $x=0$.  Evaluating Eq.~(\ref{weq}) 
and retaining both contributions results in the total conductivity:
\beq\label{eq9}
\sigma=\frac{4e^2}{\pi h}\left[\frac{\pi\kappa^2 T}{\eta}
 e^{-m/\kappa T}+\left(\frac{\pi\eta T}{3m^2}\right)^2\right]\, m\gg\kappa T.
\eeq
The exponentially-decaying term is in agreement
with the result of van Otterlo, et. al.~\cite{otterlo}, though
their coefficient is a factor of $8$ smaller.   

We now go beyond the quadratic theory we have constructed.
It is expedient to generalize to an N-component vector field, 
$\psi_\nu(\omega,\vec k)$ and introduce the quartic term
\beq
&\frac{U}{2N\beta}&\sum_{\omega_1,\omega_2,\omega_3,\omega_4}
\delta_{\omega_1+\cdots\omega_4,0} \delta_{\vec k_1+ \cdots \vec k_4,0}
\int\frac{d\vec k_1\cdots d\vec k_4}{(2\pi)^3}\nonumber\\
&&\psi_\nu(\omega_1,\vec k_1)\psi_\nu(\omega_2,\vec k_2)\dot
\psi_\mu(\omega_3,\vec k_3)\psi_\mu(\omega_4,\vec k_4)
\eeq
into our free-energy functional.
Because the phase ordering transition is no longer determined
by the vanishing of the coefficient of the quadratic term
at $k=\omega=0$, it is expedient then to redefine
$\delta$ as the bare temperature independent coefficient of the quadratic term.

We then introduce the auxiliary
field, $\lambda(\omega,\vec k)$, to decouple the quartic term by means of
a Hubbard-Stratonovich transformation.  Formally,
the auxiliary field, $\lambda(\omega,\vec k)$ is related to the correlation
function, $\langle\psi_\alpha(\omega,\vec k)\psi^*_\alpha(\omega,\vec k)\rangle$. 
Consequently,
if we invoke the saddle-point approximation, 
$\lambda(\omega,\vec k)=\sqrt{\beta}\lambda\delta_{\omega,0}\delta_{\vec k,0}$,
in the $N\rightarrow \infty$ limit,
the following self-consistency equation for the gap $m^2=\delta+ \lambda$ where
\beq\label{lambda}
\lambda=\frac{U}{\beta}\sum_{\omega_n}\int \frac{d^2k}{(2\pi)^2}
\frac{1}{m^2+k^2+\kappa^2\omega_n^2+\eta|\omega_n|},
\eeq 
must be solved.  Let us define $a=\eta/4\pi\kappa^2 T$, 
$b=\sqrt{\eta^2/4\kappa^2-m^2}/2\pi\kappa T$,
and $c=\Lambda/2\pi\kappa T$.  Upon performing
the sum over
boson frequencies and subsequent k-integration
by introducing the relativistic hard cutoff, $\Lambda$, in Eq.~(\ref{lambda}), 
we obtain 
\beq\label{m2}
m^2=\delta-\frac{U T}{2\pi}\ln\frac{\Gamma(a+ic)\Gamma(a-ic)\Lambda}
{\Gamma(a+b)\Gamma(a-b)m}
\eeq
as the self-consistency condition for the effective inverse correlation length. 
Here $\Gamma(z)$ is the Gamma function.  While Sachdev, Chubukov and
Sokol\cite{scs} have obtained a similar expression, the form derived
here is more suitable for the asymptotic analysis of the three critical
regimes of interest.

Expansion of the gamma functions when $T$ is 
the smallest parameter
reveals that the behaviour of $m^2$ depends strongly on the sign of
the quantity $\Delta=\delta+U(\Lambda+O(\eta/\kappa))/4\pi\kappa$.  
At the quantum critical point, $\Delta=0$ and consequently $m^2=0$.
At any finite temperature, the form of 
$m^2$, which enters directly into the conductivity, is determined
by the relative strength of three parameters: $\kappa T$, $\Delta$, and $\eta/\kappa$.
Consequently, the following regimes arise as depicted in Fig. (1).
\begin{figure}
\begin{center}
\epsfig{file=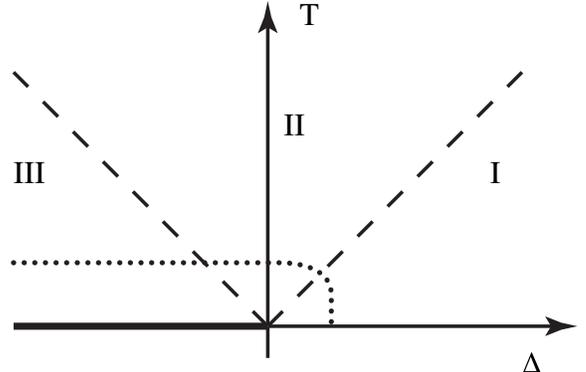, height=5.0cm}
\caption{Heuristic phase diagram as a function of temperature $T$ and 
$\Delta$.  The quantum critical point corresponds to $\Delta=0$.  Regions
I, II, and III correspond to 1) quantum disordered, 2) quantum critical,
and 3) renormalized classical.  The dotted line demarcates the region
in which dissipation dominates the physics of the correlation length.}
\label{fig1}
\end{center}
\end{figure}
 a) {\it Quantum disordered (insulator):}
Two conditions (see Fig. (1)) demarcate this region: 
$\Delta>0$ and $\Delta\gg\kappa T$.  Expansion of Eq.~(\ref{m2}) in this limit
indicates for $\Delta\gg\eta/\kappa$, $m=4\pi\kappa\Delta/U$ and
for $\eta/\kappa\gg\Delta$, 
$m=\sqrt{4\pi^2\eta\Delta/(U\ln\left(\eta U/\kappa\Delta\right))}$.
We used the fact that from microscopic considerations of the $XY$ model,
$\kappa/U\approx O(1)$ close to the transition point .
Substitution of these results into Eq.~(\ref{eq9}) leads to the corresponding
conductivity in the quantum disordered regime.  
b) {\it Renormalized Classical:}  This regime is characterized by $\Delta<0$
but $|\Delta|\gg\kappa T$ as shown in Fig. (1).  The solution to Eq.~(\ref{m2}) in this limit results
in the following forms for the gap parameter $m$:
\beq\label{level}
m=\left\{
\begin{array}{ll}
\kappa T\exp\left(-\frac{2\pi|\Delta|}{U T}\right)& |\Delta|\gg\kappa
 T\gg\eta/\kappa
\\ 
\sqrt{\eta T}\exp\left(-\frac{2\pi|\Delta|}{U T}\right) & 
|\Delta|,\eta/\kappa\gg\kappa T
\end{array}
\right. 
\eeq
In the first sub-regime, 
$m$ can be both smaller than or greater than $\eta/\kappa$.  Should $m$
be greater than $\eta/\kappa$, then the logarithmic
form for the conductivity applies.  However, because of the exponential
dependence of $m$, the resultant conductivity
\beq\label{lev}
\sigma=\frac{4 e^2}{h}\frac{\pi\kappa^2|\Delta|}{\eta U}\quad \eta/\kappa<m
\eeq
is independent of temperature. Because $\eta/\kappa$ is small, the prefactor 
of $e^2/h$ in the conductivity is large. 
This result is truly remarkable in light of the experiments by several
groups~\cite{goldman,mk,yk} that have reported a 
leveling-off of the conductivity in 2D thin film superconductors at low temperatures. 
Note that the conductivity as well as the width
of the region over which $\sigma$ is constant increase as the magnitude
of the dissipation decreases.  Hence, the model presented here is in principle
compatible with the low values of the resistivity once the leveling-off
ensues.  
Ultimately Mason and Kapitulnik~\cite{mk}
predict that dissipation leads to a ``metallic'' state
at $T=0$ in field-tuned IST's.  However, within the framework presented here,
a true ``metallic'' state
does not exist at $T=0$.  We find, in fact, that at sufficiently low
temperatures such that $m<\eta/\kappa$, the second of 
Eqs.~(\ref{level}) apply and the conductivity
\beq\label{dual}
\sigma=\left\{
\begin{array}{ll}
(e^2/h)(\eta/\kappa^2 T)\exp\left(\frac{4\pi|\Delta|}
{U T}\right)& \kappa T> \eta/\kappa
\\
(e^2/h)\exp\left(\frac{4\pi|\Delta|}{U T}\right)& \kappa T<\eta/\kappa
\end{array}
\right. 
\eeq
diverges exponentially as $T\rightarrow 0$ as is expected 
on the superconducting side. Nonetheless, the existing experiments
cannot rule out the possibility that the resistivity ultimately vanishes
at zero temperature.  

c) {\it Quantum critical region}. In this
regime, $\kappa T \gg |\Delta|$. We consider
for simplicity the transition point, $\Delta=0$.
From Eq.~(\ref{m2}), it follows that
 $m=\sqrt{2\pi\eta T} f_1(\kappa^2 T/\eta)$, where $f_1$ is a
numerically calculable function. We find that in the
limiting case $\kappa T \ll \eta/\kappa$, $m=\sqrt{2\pi A\eta T}$, where
$A=2.68723$ is the larger of the roots of the equation 
$\Gamma(x)\sqrt{x}=\sqrt{2\pi}$. In the opposite limit, $\kappa T \gg
\eta/\kappa$, we recover the well-known~\cite{chubukov,scs} result 
that $m=\Theta\kappa T$ with
$\Theta = 2\ln((\sqrt{5}+1)/2)$. In general, the 
conductivity is given by 
$\sigma=(e^2/h) f_2(\kappa^2 T/\eta, \kappa/U)$,
with $f_2$ a universal function. To calculate
$f_2$, we investigate Eq.~(\ref{weq}). 
For $\kappa T \gg \eta/\kappa$, the main
contribution to the integral over $x$ comes from the vicinity of $x_{\pm}$
leading to a conductivity of the form, $\sigma=2K_1 (e^2/h)(\kappa^2 T/\eta)$
with
\beq
K_1=\int_0^\infty \frac{t^3 dt}{(t^2+\Theta^2/4)\sinh^2(t^2+\Theta^2/4)}=0.0149.
\eeq
For  $\kappa T \ll \eta/\kappa$, it is convenient to perform the 
k-integration in Eq.~(\ref{weq})) first. We obtain for the conductivity,
$\sigma=(2K_2/\pi)(e^2/h)$ where 
\beq
K_2=\int_0^\infty \frac{dt}{\sinh^2 t} \left[ 1-\frac{\pi A}{t}
\left( \frac{\pi}{2}-\arctan\frac{\pi A}{t} \right) \right]=0.0076.\nonumber
\eeq
We see that in the limit $T\rightarrow 0$, the 
conductivity acquires a universal value independent of 
the dissipation.  The independence of the critical resistance on $\eta$
is a consequence of hyperuniversality~\cite{otterlo}.  One should note,
though, that the extremely small numerical prefactor, is likely to be an artifact
of this model. 

Our analysis of the role of dissipation in the vicinity of the 
quantum critical point 
has revealed a key new feature,
namely the leveling of the resistivity at sufficiently low temperatures.
To determine if this leveling is experimentally relevant,
we evaluate the resistivity by inverting Eq. (\ref{lev}). As shown
previously\cite{wag}, $U/\kappa=7/2$ and $\kappa=1/4E_c=0.3K^{-1}$\cite{mooij}
 close to
the quantum critical point. Here $E_c$ is the charging energy.  
Consequently, the level resistivity reduces to 
$R=R_o(7/2\pi)(\eta/\kappa)/\|\Delta|$, where $R_o=h/4e^2=6.4k\Omega$.  
Typical values for the leveled resistance are $R=10\Omega$\cite{mooij}
which would require the ratio $\eta/\kappa|\Delta|=.01$. For
$|\Delta|$ in the range $[0.01,.1]$, we find that the corresponding
value of the dissipation ranges between $10^{-5}<\eta/\kappa<10^{-4}$ which
is certainly in the weak dissipation regime as required by
the present analysis.  From the condition $m\approx\eta/\kappa$,
we find from Eq. (\ref{level}) that the 
leveling behavior ceases at $T\sim O(10mK)$. Hence, we find that for realistic
parameters, our analysis predicts a level resistance that is consistent
with experimental trends.

\acknowledgements
This work was funded by the DMR
of the NSF and the ACS petroleum research fund.

\end{document}